\documentclass[aps,pra,reprint,superscriptaddress,amsmath,amssymb,amsfonts,longbibliography]{revtex4-1}

\usepackage{graphicx}
\usepackage{hyperref}
\hypersetup{colorlinks=true,linkcolor = blue, urlcolor = blue, citecolor = blue}
\usepackage{xcolor}
\usepackage{stix}

\usepackage{mathtools}
\DeclarePairedDelimiter\myKet{|}{\rcurvyangle} 
\DeclarePairedDelimiter\myBra{\lcurvyangle}{|}	
\usepackage{upgreek}
\newcommand{\aDagg}[1]{a^{\dagger}_{#1}} 
\newcommand{\ket}[1]{\left| #1 \right\rangle} 
\newcommand{\bra}[1]{\left\langle #1 \right|} 
\DeclareMathOperator{\Tr}{Tr} 
\newcommand{\Tau}{\mathcal{T}} 
\newcommand{\diffD}{\mathrm{d}} 
\newcommand{\kB}{k_{\mathrm{B}}} 
\DeclareMathOperator{\csch}{csch} 
\newcommand{\matel}[2]{
\ifnum 0=#1\relax
S_{#2}
\else
S_{#2}^{*}
\fi
}

\newcommand{\norm}[1]{\lVert#1 \hspace{1mm}\rVert}
\newcommand{\telF}{\mathcal{F}_{\mathrm{tel}}} 

\newcommand{\vecB}[1]{\mathbf{#1}} 
\newcommand{\rmA}{\mathrm{A}}
\newcommand{\rmB}{\mathrm{B}}

\begin{document}


\title{Quantum teleportation of single-electron states}


\author{Edvin Olofsson}
\affiliation{Physics Department and NanoLund, Lund University, Box 118, 22100 Lund, Sweden}

\author{Peter Samuelsson}
\affiliation{Physics Department and NanoLund, Lund University, Box 118, 22100 Lund, Sweden}

\author{Nicolas Brunner}
\affiliation{D\'{e}partement de Physique Appliqu\'{e}e, Universit\'{e} de Gen\`{e}ve, 1211 Gen\`{e}ve, Switzerland}

\author{Patrick P. Potts}
\affiliation{Physics Department and NanoLund, Lund University, Box 118, 22100 Lund, Sweden}


\date{\today}

\begin{abstract}
We consider a scheme for on-demand teleportation of a dual-rail electron qubit state, based on single-electron sources and detectors. The scheme has a maximal efficiency of 25\%, which is limited both by the shared entangled state as well as the Bell-state measurement. We consider two experimental implementations, realizable with current technology. The first relies on surface acoustic waves, where all the ingredients are readily available. The second is based on Lorentzian voltage pulses in quantum Hall edge channels. As single-electron detection is not yet experimentally established in these systems, we consider a tomographic detection of teleportation using current correlators up to (and including) third order. For both implementations we take into account environmental effects. 
\end{abstract}


\maketitle

\section{Introduction}
Quantum teleportation was introduced by Bennett et al.\ in 1993~\cite{Bennett93} and the first experimental implementations, using photon polarizations, started to appear in the late 90s~\cite{Bouwmeester,RomeTel}. Implementations in other photonic and matter based systems have since followed \cite{SingleRail, TimeBinPhot,TrappedIons,NMR,SupCond}. Recent developments include ground-to-satellite \cite{ren:2017} and chip-to-chip \cite{llewellyn:2020} teleportation, quantum secret sharing \cite{lee:2020}, and teleportation in high dimensions \cite{luo:2019}. The objective of quantum teleportation is the transfer of a quantum state between two parties, Alice and Bob, using classical communication, shared entanglement and local measurements. In addition to directly transferring quantum states, it may serve as a part of an entanglement swapping scheme \cite{PhysRevLett.71.4287,pan:1998}. Teleportation has important applications in quantum communication, quantum computing, and quantum networks \cite{NaturePhotonics,grosshans:2016,zeilinger:2018}, enabling important building blocks such as quantum repeaters \cite{dur:1998, sangouard:2011} and state transfer between propagating photons and solid-state quantum memories \cite{QuDotMem,RareEarthMem}. In addition to practical applications, teleportation experiments provide a simple way to demonstrate control over quantum systems, since it involves some of the fundamental building blocks of quantum information: state preparation, entanglement generation, measurements, and unitary transformations conditioned on measurement outcomes.

Relying on the mature semiconducting technology, electronic circuits provide a promising avenue for quantum technologies due to their potential for scalability and integration with existing devices. Motivated by recent progress in the generation and manipulation of single-electron states in mesoscopic systems, a field referred to as electron quantum optics \cite{BaeuerleRev, LevitonExp, NJP, SolidiB, SAWBS,doi:10.1002/andp.201300181}, we propose a scheme for performing quantum teleportation of single-electron states. The scheme is based on electronic analogs of optical components such as beamsplitters and phase shifters to manipulate electrons in a dual-rail qubit configuration, which consists of two spatial modes that an electron can occupy. The (arbitrary) qubit state that is going to be teleported is prepared by a single-electron source combined with a tunable beamsplitter and a tunable phase shifter. The entanglement required for teleportation is generated by two single-electron sources together with a pair of 50/50 beamsplitters \cite{NJP, SolidiB} and the Bell-state measurement is implemented using beamsplitters and charge detectors. 
In this scheme, the efficiency of teleportation is restricted for two reasons: First, particle-number superselection renders the entangled state useless in 50\% of the cases. Second, due to the linearity of the system, the Bell-state measurement has a finite success rate of 50\% \cite{vaidman:1999,lutkenhaus:1999,calsamiglia:2001} so that the teleportation succeeds with an overall probability of 25\%. We further note that the efficiency is reduced by another 50\% if the final unitary transformation is not applied.

We consider two possible experimental implementations. The first one is based on using surface acoustic waves (SAW) to transport single electrons between static quantum dots, which can act as detectors \cite{SAWSurf,BaeuerleRev}.  Since maintaining coherence is a challenge in this type of system \cite{BaeuerleRev,SAWBS}, we consider dephasing due to fluctuating electric fields.
The second implementation is based on levitons traveling in chiral edge states that occur in the quantum Hall regime. A leviton is a single-electron excitation on top of the Fermi sea created by applying a Lorentzian shaped voltage pulse to a metallic contact \cite{LevitovMatPhys,Levitov, LevitonExp}. Despite promising recent efforts \cite{QDotDet,glattli:2020}, single-electron detection has not been demonstrated yet for this type of setup. For this reason, we will theoretically demonstrate how to perform state tomography of Bob's post measurement state by periodically repeating the experiment and measuring zero frequency currents and current cross-correlators up to (and including) order three. A rigorous connection to the observables of the idealized single-shot scenario is found at zero temperature. We also consider how finite temperatures alter the results. In both of the experimental implementations we find that the effect of the environment is to introduce noise that can be described as phase damping~\cite{MikeAndIke}.

Chiral edge channels have been considered before for quantum teleportation \cite{Beenakker}. While we propose performing full state tomography on Bob's post-measurement state, in Ref.~\cite{Beenakker} the teleportation is demonstrated by simultaneously teleporting a hole, in a scheme analogous to entanglement swapping \cite{PhysRevLett.71.4287}, and verifying that the resulting electron-hole pair is entangled by measuring low frequency current correlators. Furthermore, our scheme relies on single-electron sources to provide teleportation \textit{on demand}.
Other teleportation schemes in solid state systems include teleporation of electron spins~\cite{PhysRevB.69.035332,PhysRevLett.96.246801,qiao:2019}, transmon qubits~\cite{SupCond}, nitrogen-vacancy center qubits~\cite{Pfaff532}, and teleportation from photons to solid-state quantum memories~\cite{QuDotMem,RareEarthMem}. Finally, we note that our results are in complete agreement with a simultaneous and independent work \cite{FriisTel}.

The rest of the paper is structured as follows: In Sec.\ \ref{sec:IdealSetup} the teleportation scheme is introduced, and we derive results for the teleportation efficiency. Section \ref{sec:ExperimentalImplementations} contains the considered experimental implementations, where Secs.~\ref{sec:SAW} and \ref{sec:Levitons} are devoted to the SAW and leviton implementations, respectively. We conclude the article in Sec.~\ref{sec:conclusions}.


\section{Teleportation with electrons}\label{sec:IdealSetup}
Here we present a scheme for performing quantum teleportation with dual-rail electron qubits, provided by two orthonormal spatial modes. The scenario considered in this section relies on single-electron emission and detection. The setup is presented in Fig.~\ref{fig:telSetup}~(b) and consists of four parts. The region denoted \textit{state preparation} prepares the state that is going to be teleported from Alice to Bob. The \textit{entanglement generation} region is used to generate a state that is entangled between Alice and Bob. Alice performs a measurement as a part of the teleportation protocol in the \textit{electron detection} region. In the \textit{state tomography} region Bob performs state tomography on the state that he received to determine how well the protocol worked.
The scattering matrix describing the complete setup is given in App.~\ref{App:Full}.

Before turning to teleportation with electrons, we briefly recapitulate the original scheme for teleportation \cite{Bennett93}, presented schematically in Fig.~\ref{fig:telSetup}\,(a). Alice receives the state $\ket{\psi}$ that is to be teleported. Additionally, she and Bob are each in possession of one half of an entangled state $\ket{\phi}$, usually taken as one of the four Bell states. Alice performs a combined measurement in the Bell basis on $\ket{\psi}$ and her part of the entangled state. After the measurement, Bob's part of $\ket{\phi}$ will be left in the state $U\ket{\psi}$, where $U$ is a unitary transformation determined by the outcome of Alice's measurement. Alice then communicates the outcome of her measurement to Bob, who can use the information to apply the inverse transformation $U^{\dagger}$ to his post-measurement state. Bob is then left with the state $\ket{\psi}$ and the protocol is completed.

\subsection{State preparation and entanglement generation}\label{sec:StatePrep}

\begin{figure*}
\includegraphics[width=\textwidth]{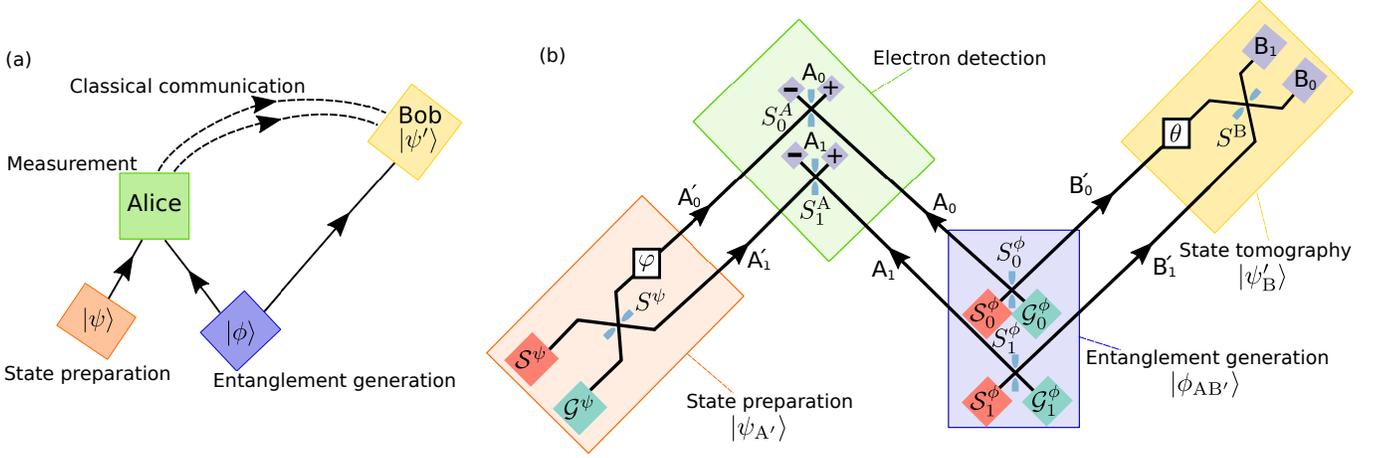}
\caption{\label{fig:telSetup}Schematic view of the original teleportation protocol (a) and our proposed setup for single-electron teleportation (b).  (a) Alice and Bob are each given one half of an entangled state $\ket{\phi}$. Alice also receives an unknown qubit state $\ket{\psi}$. She performs a measurement on the combined state of the qubit and her part of the entangled state, and sends the result to Bob.  Based on the measurement outcome, Bob performs a unitary operation on his part of the entangled state to recover the unknown qubit from the post-measurement state $\ket{\psi^{\prime}}$. (b) The $\mathcal{S}_i$ denote single-electron sources, $\mathcal{G}_i$ empty inputs (grounded contacts in the implementation based on levitons). Beamsplitters are labeled by $S$ and phase-shifters by $\varphi$ and $\theta$. The A and A$^{\prime}$ modes propagate to Alice while the B and B$^{\prime}$ propagate to Bob. Electrons are detected by Alice at four detectors at $\mathrm{A}^{\pm}_0$ and at $\mathrm{A}^{\pm}_1$. The aim of the experiment is to transfer a superposition of A$^{\prime}$ modes to a superposition of B$^{\prime}$ modes. Bob can perform state tomography on his part of the post-measurement state in order to verify  teleportation. He selects which component of the Bloch vector to measure by adjusting $S^{\mathrm{B}}$ and $\theta$.}
\end{figure*}

The first step in the protocol is the preparation of the state that is going to be teleported as well as the shared entangled state. To this end, three single-electron sources, denoted by $\mathcal{S}^{\psi}$ and $\mathcal{S}^{\phi}_j$, $j=0,1$, emit single electrons, which will travel towards a first set of beamsplitters, $S^{\psi}$ and $S^{\phi}_j$. This can be described by the state 
\begin{equation}
\ket{\Psi} = \aDagg{\mathcal{S}^{\psi}}\aDagg{\mathcal{S}^{\phi}_0}\aDagg{\mathcal{S}^{\phi}_1}\ket{\Omega}.
\end{equation}
Here the $\aDagg{i}$ are fermionic creation operators that populate mode $i$ with an electron, and $\ket{\Omega}$ denotes the vacuum or the Fermi sea, see below.
The beamsplitters are  described by the scattering matrices
\begin{align}\label{eq:stateScattMat}
\begin{split}
S^{\psi} &= 
\begin{pmatrix}
i\sqrt{R}e^{-i\varphi} & \sqrt{D}e^{-i\varphi}\\
\sqrt{D} & i\sqrt{R}\\
\end{pmatrix}
\end{split}
\end{align}
and
\begin{align}\label{eq:entangleScattMat}
\begin{split}
S_{j}^{\phi}=\frac{1}{\sqrt{2}} 
\begin{pmatrix}
i & 1\\
1 & i
\end{pmatrix},
\end{split}
\end{align}  
where $R$ is the reflection probability and $D$ the transmission probability, with $R+D=1$, and $\varphi$ is a phase difference. 
 Using Eqs.\ \eqref{eq:stateScattMat} and \eqref{eq:entangleScattMat}, $\ket{\Psi}$ can be expressed as
\begin{align}\label{eq:middleState}
\begin{split}
\ket{\Psi} = &\frac{1}{2}\left(i\sqrt{R}e^{-i\varphi}a^{\dagger}_{\mathrm{A}_{0}^{\prime}} + \sqrt{D}a^{\dagger}_{\mathrm{A}_{1}^{\prime}}\right)\left(ia^{\dagger}_{\mathrm{A}_{0}} + a^{\dagger}_{\mathrm{B}_0^{\prime}}\right)\\ 
&\times\left(ia^{\dagger}_{\mathrm{A}_{1}} + a^{\dagger}_{\mathrm{B}_1^{\prime}}\right) \ket{\Omega}.
\end{split}
\end{align}
The A$_0$ and A$_1$ modes represent electrons traveling from the $\mathcal{S}_0^{\phi}$ and $\mathcal{S}_1^{\phi}$ sources, respectively, to Alice's detector regions. A$^{\prime}_0$ and A$^{\prime}_{1}$ modes instead represent electrons created at $\mathcal{S}^{\psi}$. The B$_0^{\prime}$ and B$_1^{\prime}$ modes correspond to electrons traveling to Bob. Each mode is illustrated in Fig.\ \ref{fig:telSetup}~(b).

It is instructive to consider the state that is to be teleported and the shared entangled state individually. 
Tracing over the A and B$^{\prime}$ modes yields
\begin{equation}
\Tr_{\mathrm{A}\mathrm{B}^{\prime}} (\ket{\Psi}\bra{\Psi}) =  \ket{\psi_{\mathrm{A}^{\prime}}}\bra{\psi_{\mathrm{A}^{\prime}}},
\end{equation}
where 
\begin{equation}
\ket{\psi_{\mathrm{A}^{\prime}}}=\left(i\sqrt{R}e^{-i\varphi}a^{\dagger}_{\mathrm{A}_{0}^{\prime}} + \sqrt{D}a^{\dagger}_{\mathrm{A}_{1}^{\prime}}\right)\ket{\Omega_{\mathrm{A}^{\prime}}}
\end{equation} 
is the dual-rail qubit that we wish to teleport. Here $\ket{\Omega_{\mathrm{A}^{\prime}}}$ refers to the vacuum state associated with the A$^{\prime}$ modes.
If we instead trace over the A$^{\prime}$ modes, the result is
\begin{equation}
\Tr_{\mathrm{A}^{\prime}} (\ket{\Psi}\bra{\Psi})  = \ket{\phi_{\mathrm{AB}^{\prime}}}\bra{\phi_{\mathrm{AB}^{\prime}}},
\end{equation}
where
\begin{equation}\label{eq:TraceAPrime}
\ket{\phi_{\mathrm{AB}^{\prime}}} =\frac{1}{2}\left(ia^{\dagger}_{\mathrm{A}_{0}} + a^{\dagger}_{\mathrm{B}_0^{\prime}}\right)\left(ia^{\dagger}_{\mathrm{A}_{1}} + a^{\dagger}_{\mathrm{B}_1^{\prime}}\right) \ket{\Omega_{\mathrm{AB}^{\prime}}},
\end{equation}
where $\ket{\Omega_{\mathrm{AB}^{\prime}}}$ is the vacuum state associated with the A and B$^{\prime}$ modes.
The above state is entangled between Alice and Bob and can be used to violate a Bell inequality~\cite{NJP,SolidiB}. However, the parts of $\ket{\phi_{\mathrm{AB}^{\prime}}}$ that correspond to Bob having zero or two electrons cannot be used for teleportation in our setup, where the particle number is conserved. In those cases, Bob does not receive a dual-rail qubit. From Eq.\ \eqref{eq:TraceAPrime} we see that this reduces the success probability for teleportation by 50\%.

Since we  consider teleportation of dual-rail qubit states, it is instructive to consider the subspace spanned by these states (i.e., the subspace where there is exactly one electron in modes A, A$^{\prime}$, and  B$^{\prime}$ respectively). States projected onto this subspace will be denoted by $\myKet{}$. Since the state to be teleported already is a dual-rail qubit state, we find $\myKet{\psi_{\mathrm{A}^{\prime}}} = \ket{\psi_{\mathrm{A}^{\prime}}}$.
Introducing $\aDagg{\mathrm{A}_k}\aDagg{\mathrm{B}^{\prime}_l}\ket{\Omega} = \myKet{k_{\mathrm{A}},l_{\mathrm{B}^{\prime}}}$
we find
\begin{equation}
\myKet{\phi_{\mathrm{AB}^{\prime}}} = \frac{i}{2}\left(\myKet{0_{\mathrm{A}},1_{\mathrm{B}^{\prime}}}-\myKet{1_{\mathrm{A}},0_{\mathrm{B}^{\prime}}}\right) = \frac{i}{\sqrt{2}}\myKet{\Uppsi^-_{\mathrm{AB}^{\prime}}},
\end{equation}
where $\myKet{\Uppsi^-_{\mathrm{AB}^{\prime}}}$ denotes one of the Bell states $\lbrace \myKet{\Uppsi^\pm_{\mathrm{AB}^{\prime}}},\myKet{\Upphi^\pm_{\mathrm{AB}^{\prime}}}\rbrace$, see App.~\ref{App:ProjectionsSQ} for expressions in terms of creation operators.
The total state is therefore projected onto
\begin{align}\label{eq:totalProjectedState}
\begin{split}
\myKet{\Psi} &= \frac{i}{\sqrt{2}} \myKet{\psi_{\mathrm{A}^{\prime}},\Uppsi^-_{\mathrm{AB}^{\prime}}} = \frac{-i}{2\sqrt{2}}\Bigl( \myKet{\Uppsi^-_{\mathrm{A}^{\prime}\mathrm{A}},\psi_{\mathrm{B}^{\prime}}} \\&+ \myKet{\Uppsi^+_{\mathrm{A}^{\prime}\mathrm{A}},\sigma_z\psi_{\mathrm{B}^{\prime}}}
- \myKet{\Upphi^+_{\mathrm{A}^{\prime}\mathrm{A}}, \sigma_x\psi_{\mathrm{B}^{\prime}}} + \myKet{\Upphi^-_{\mathrm{A}^{\prime}\mathrm{A}}, i\sigma_y\psi_{\mathrm{B}^{\prime}}} \Bigr).
\end{split}
\end{align}
Here $\myKet{\Uppsi^+_{\mathrm{A}^{\prime}\mathrm{A}}, \sigma_z \psi_{\mathrm{B}^{\prime}}
}$ denotes a state where the dual-rail qubits at A$^{\prime}$ and A encode the Bell state $\myKet{\Uppsi^+_{\mathrm{A}^{\prime}\mathrm{A}}}$ and the qubit at B$^{\prime}$ is described by the state $\sigma_z\myKet{\psi_{\mathrm{B}^{\prime}}}$, with $\sigma_j$, $j=x,y,z$, denoting the Pauli matrices. Expressions in terms of creation operators are given in App.~\ref{App:ProjectionsSQ}. The state $\myKet{\Psi}$ is (up to the normalization) equivalent to the pre-measurement state in conventional quantum teleportation schemes~\cite{Bennett93, MikeAndIke}. We note that mapping fermionic states onto qubit states can be problematic~\cite{FriisPartTrace,ding:2020}, in particular when taking partial traces. Here we  perform all calculations using fermionic states, and merely use the qubit notation (e.g.\ $\myKet{0_{\rmA},1_{\rmB^{\prime}}}$) for illustrative purposes.

\subsection{Electron detection}\label{sec:electronDetection}
Next, Alice performs her measurement. To this end, any electron traveling to Alice passes through another set of beamsplitters, $S^{A}_0$ and $S^{A}_1$. These are also described by the scattering matrix defined in Eq.\ \eqref{eq:entangleScattMat}. Then, Alice performs single-electron detection, determining the number of electrons in each mode A$_j^{\pm}$. The state prior to Alice's measurement can be written as 
\begin{equation}\label{eq:finalState}
\ket{\Psi} = \frac{1}{2}\ket{T} + \frac{\sqrt{3}}{2}\ket{R},
\end{equation}
where we have introduced the normalized and orthogonal states $\ket{T}$ and $\ket{R}$. In terms of creation operators, $\ket{T}$ has the form
\begin{align}\label{eq:T}
\begin{split}
&\ket{T}=\frac{1}{2}\biggl[\left(\aDagg{\mathrm{A}^+_0}\aDagg{\mathrm{A}^+_1}+ \aDagg{\mathrm{A}^-_0}\aDagg{\mathrm{A}^-_1}\right)\left(i\sqrt{R}e^{-i\varphi}\aDagg{\mathrm{B}_0^{\prime}} +  \sqrt{D}\aDagg{\mathrm{B}_1^{\prime}}\right)  \\
&-i\left(\aDagg{\mathrm{A}^+_0}\aDagg{\mathrm{A}^-_1}- \aDagg{\mathrm{A}^-_0}\aDagg{\mathrm{A}^+_1}\right)\left(i\sqrt{R}e^{-i\varphi}\aDagg{\mathrm{B}_0^{\prime}} -\sqrt{D}\aDagg{\mathrm{B}_1^{\prime}}\right) \biggr] \ket{\Omega},
\end{split}
\end{align}
 and it is the part of $\ket{\Psi}$ that is useful for teleportation in the considered scenario. In terms of the B$_0^{\prime}$ and B$_1^{\prime}$ modes, the terms in $\ket{T}$ have a similar structure to $\ket{\psi}$.
The terms in $\ket{R}$ correspond to cases where Bob cannot receive a coherent dual-rail qubit state after Alice's measurement. This may happen, if Bob receives zero or two electrons, or if Alice detects both electrons at A$_0$ (A$_1$) such that the remaining electron necessarily is at B$^{\prime}_1$ (B$^{\prime}_0$). The form of $\ket{R}$ is discussed further in App.~\ref{App:Full}. We note that $\ket{T} \neq \myKet{\Psi}$ since not all terms in $\myKet{\Psi}$ are useful for teleportation. This is due to a limitation in the measurement discussed below.  
We note that lifting the constraint of particle-number superselection, $\ket{R}$ may also be useful for teleportation \cite{FriisTel}.

The electron detection can be described by a positive operator valued measure~(POVM) with elements $\lbrace E(X)\rbrace$ associated to the set of measurement outcomes $\lbrace X\rbrace$.  
Bob's post-measurement state $\rho_{\mathrm{B}}(X)$ can then be found by taking the partial trace  over the A$_j^{\pm}$ modes
\begin{equation}\label{eq:postMeasPOVM}
\rho_{\mathrm{B}}(X)=\frac{1}{p(X)}\Tr_{\mathrm{A}^{\pm}}\left(E(X)\rho\right),
\end{equation} 
where $\rho = \ket{\Psi}\bra{\Psi}$ and $p(X) = \bra{\Psi} E(X)\ket{\Psi} $ is the probability that the result of Alice's measurement is $X$. 

Eq.\ \eqref{eq:T} suggests that there are four outcomes of Alice's measurement for which Bob's post-measurement state is related to the original input qubit via a unitary transformation.
We will therefore focus on these outcomes, writing $ X=(s_0,s_1)$, $s_i\in \lbrace +,-\rbrace$, for the outcomes where one electron is detected at A$_0^{s_0}$ and one at A$_1^{s_1}$. App.~\ref{App:POVM} contains the definition of the full POVM for Alice's measurement.

 Applying Eq.\ \eqref{eq:postMeasPOVM} for the considered outcomes yields
\begin{align}\label{eq:BobPostMeas}
\begin{split}
&\rho_{\mathrm{B}}(+,+) = \rho_{\mathrm{B}}(-,-) = \ket{\psi} \bra{\psi}, \\
&\rho_{\mathrm{B}}(+,-) = \rho_{\mathrm{B}}(-,+) = \sigma_z\ket{\psi} \bra{\psi}\sigma_z,
\end{split}
\end{align}
where $\sigma_z = N_{\rmB^{\prime}_0}-N_{\rmB^{\prime}_1}$ 
and $N_{k} = \aDagg{k}a_k$. 
Each of these outcomes occurs with a probability of 1/16. This means that the teleportation scheme is successful 25\% of the time,  if a feed-forward mechanism is implemented to apply $\sigma_z$ when Alice measures +$-$ or $-$+. This unitary can be implemented  by introducing a $\pi$ phase shift between the B$_0^{\prime}$ and B$_1^{\prime}$ modes. Without an active feed-forward, the efficiency drops to 12.5\%, which corresponds to the ++ and $--$ outcomes of Alice's measurement. 

 The  protocol therefore requires Alice to send one of three messages, which can be communicated using two classical bits, to Bob. When she measures ++ or $--$ Bob should do nothing to his state, if she measures +$-$ or $-$+ he should apply the phase shift. For all other outcomes the protocol failed. In the absence of the feed-forward, it is sufficient to send a single classical bit to communicate if teleportation was successful (outcomes ++ and $--$) or not. The feasibility of implementing the feed-forward in practice will depend on the specifics of the experimental implementation. Of the two implementations that will be discussed in Sec.~\ref{sec:ExperimentalImplementations}, the SAW seems more suited for this because the feed-forward requires single-electron detection.

To connect to the standard teleportation protocol, we consider how the electron detection looks in the dual-rail qubit subspace. In the standard teleportation protocol, Alice measures her qubits in the Bell basis and sends the result to Bob. In the dual-rail qubit subspace, the electron detection corresponds to the measurement basis $\myKet{\Uppsi^-_{\mathrm{A}^{\prime}\mathrm{A}}}$ (outcomes ++, $--$), $\myKet{\Uppsi^+_{\mathrm{A}^{\prime}\mathrm{A}}}$ (outcomes +$-$, $-$+), and $1/\sqrt{2}(\myKet{\Upphi^+_{\mathrm{A}^{\prime}\mathrm{A}}}\pm\myKet{\Upphi^-_{\mathrm{A}^{\prime}\mathrm{A}}})$ (2 electrons at A$_0$ or A$_1$), see Apps.~\ref{App:ProjectionsSQ} and \ref{App:POVM}. Therefore, Alice's measurement cannot distinguish between  $\myKet{\Upphi^+_{\mathrm{A}^{\prime}\mathrm{A}}}$ and $\myKet{\Upphi^-_{\mathrm{A}^{\prime}\mathrm{A}}}$, and the Bell state measurement is incomplete. Consequently, the measurement will only result in a useful outcome half of the time. This restriction is general for linear systems \cite{vaidman:1999,lutkenhaus:1999,calsamiglia:2001}, but can be overcome using additional entangled degrees of freedom \cite{kwiat:1998,beenakker:2004}. Combined with the 50\% chance to find the state in the dual-rail qubit subspace, we arrive at an overall efficiency of the protocol of 25\% (including feed-forward), in agreement with the discussion above.

\subsection{State tomography}\label{sec:StateTomography}
To verify that teleportation occurred, Bob can perform quantum state tomography on his part of the state after Alice has performed her measurement.  The Bloch vector $\vecB{r}$ that describes $\ket{\psi}$ has components $r_j = \bra{\psi}\sigma_j\ket{\psi}$, where $\sigma_j$ denotes the $j$ Pauli matrix in the dual-rail qubit space (see App.~\ref{App:ProjectionsSQ} for expressions in terms of creation operators). The Bloch vector of Bob's post measurement state is given by $r_j^{\prime} = \Tr(\sigma_j\rho_{\mathrm{B}})$, where we have here chosen to focus on the $++$ outcome, $\rho_{\mathrm{B}} = \rho_{\mathrm{B}}(+,+)$. The Bloch vector component $r_j$ can be measured by determining the occupation of the modes B$_0$, B$_1$ after an additional beamsplitter and phase shift $\theta$, described by the scattering matrix
\begin{equation}
S^{\mathrm{B}}
=
\begin{pmatrix}
\sqrt{D^{\prime}}e^{-i\theta} & -i\sqrt{R^{\prime}}\\
-i\sqrt{R^{\prime}}e^{-i\theta} & \sqrt{D^{\prime}}
\end{pmatrix},
\end{equation}
where $D^{\prime}$ and $R^{\prime}$ are the transmission and reflection probabilities. A sketch of the tomography setup is shown in Fig.\ \ref{fig:telSetup}~(b). The settings for $S^{\mathrm{B}}$ that are needed to perform all three measurements required for state tomography are provided in Table~\ref{tab:tomoSettings}.

\begin{table}[b!]
\begin{ruledtabular}
\begin{tabular}{ccc}
Measurement & $D^{\prime}$ & $\theta$\\
\hline
$r^{\prime}_x$ & 1/2 & $\pi$/2\\
$r^{\prime}_y$ & 1/2 & 0\\
$r^{\prime}_z$ & 1 & 0\\
\end{tabular}
\end{ruledtabular}
\caption{The phase and beamsplitter settings required for the different state tomography measurements.}\label{tab:tomoSettings}
\end{table}

It is illustrative to give $\vecB{r}^{\prime}$ in terms of the POVM for Alice's electron detection
\begin{align}\label{eq:TomoExpr}
\begin{split}
{r}^{\prime}_i &= \frac{\bra{\Psi_i}E(+,+)\sigma_i\ket{\Psi_i}}{p(+,+)} \\
&= \frac{\langle N_{\mathrm{A}_0^{+}}N_{\mathrm{A}_1^{+}}(N_{\mathrm{B}_0}-N_{\mathrm{B}_1})\rangle_i}{\langle N_{\mathrm{A}_0^{+}} N_{\mathrm{A}_1^{+}}(I-N_{\mathrm{A}_0^{-}} + N_{\mathrm{A}_1^{-}})\rangle},
\end{split}
\end{align}
where the expectation values are with respect to the full state of the system (prior to Alice's measurement) and the subscript $i$ denotes that $S^{\mathrm{B}}$ uses the settings for measuring the $i$ component of the Bloch vector $\vecB{r}^{\prime}$. Equation~\eqref{eq:TomoExpr} shows how state tomography can be performed by measuring occupation numbers. Since the teleportation setup contains only three electrons, terms that contain more than three number operators will not contribute and have been dropped in the final expression.

\section{Experimental implementations}\label{sec:ExperimentalImplementations}
The picture presented so far is idealized in the sense that environmental effects are neglected and that it assumes that single-electron detection is readily available. We will now present two possible experimental implementations and study how the picture changes when environmental effects are included. The first implementation uses quantum dots to isolate single electrons and then SAWs to transport them. Single-electron detection is available for this approach \cite{SAWSurf}. We will consider environmental effects by including random fluctuations in the electric field surrounding the itinerant electrons. In the second implementation we consider levitons in the chiral edge states of the integer quantum Hall effect. Work towards single-electron detection in chiral edge states has been performed, but it has not been achieved yet \cite{QDotDet,glattli:2020}. This prevents the implementation of single-shot teleportation. However, ensemble averages for a given outcome of Alice's measurement are available by measuring low-frequency current correlators. We consider the influence of finite temperatures in this implementation.  
 
\subsection{Surface acoustic waves}\label{sec:SAW}
SAWs are sound waves that travel on the surface of a material. If the material in question is piezoelectric, the SAW will create a co-moving disturbance in the electric potential that can be used as a moving quantum dot. If a single electron is trapped in a quantum dot and a SAW train is launched at the dot, the electron can be transferred from the static to the moving dot. It is then possible to transfer single electrons between static quantum dots by connecting them with a SAW channel \cite{SAWSurf,SAWTransf,SAWTransf2}. The electron traveling with the SAW can be captured in a static dot, where its presence can be detected \cite{SAWSurf}. The possibility of performing single-electron detection is an advantage of the SAW approach. A beamsplitter for SAW-based single-electron devices has been demonstrated recently \cite{SAWBS} and phase shifters could be implemented using side-gates. Realizing phase coherent transport of single-electrons using SAWs is an ongoing research topic \cite{BaeuerleRev}. Including decoherence is thus important to describe a realistic implementation of the teleportation scheme. To describe decoherence we use a phase averaging procedure, where in each arm of the teleportation setup we introduce a fluctuating phase representing fluctuating voltages felt by the scattering electrons~\cite{PhysRevLett.92.056805,PhysRevB.70.125305}. Alternative approaches to decoherence are provided by dephasing and voltage probes~\cite{PhysRevLett.97.066801,PhysRevB.75.035340}.

As in Sec.\ \ref{sec:StateTomography}, we will focus only on the ++ outcome of Alice's measurement, since this is sufficient to demonstrate the teleportation protocol. Since fluctuations in the electric field alter the energy of the propagating electrons, the phase accumulated while propagating in the field will also fluctuate.
We therefore model the effect of fluctuating electric fields by random phases that are introduced for the A$_{0,1}$, A$^{\prime}_{0,1}$ and B$^{\prime}_{0,1}$ modes, see Fig.\ \ref{fig:telSetup}~(b). We assume that the voltages are constant during each run of the experiment. For a fixed value of the fluctuating phases, Bob's post-measured state reads
\begin{equation}
\rho_{\mathrm{B}}(\varphi^{\prime}) =
\begin{pmatrix}
R & i\sqrt{RD}e^{-i(\varphi+\varphi^{\prime})}\\
-i\sqrt{RD}e^{i(\varphi+\varphi^{\prime})} & D
\end{pmatrix},
\end{equation}
where 
\begin{equation}
\varphi^{\prime} = \varphi_{\rmA^{\prime}_0}+\varphi_{\rmA_1}+\varphi_{\rmB_0}-(\varphi_{\rmA^{\prime}_1}+\varphi_{\rmA_0}+\varphi_{\rmB_1}).
\end{equation}
We now assume that the individual phases $\varphi_j$ follow Gaussian distributions with vanishing average and variance $\sigma_j^2$. 
The total phase $\varphi^{\prime}$ will be a Gaussian random variable with vanishing average, and  variance $\sigma^2 = \sum_j \sigma_j^2$. Bob's post-measurement state is given by an average over the random phases
\begin{align}
\begin{split}
\rho_{\mathrm{B}} &=\frac{1}{\sqrt{2\pi\sigma^2}}\int_{-\infty}^{\infty} \diffD \varphi^{\prime}e^{-{\varphi^{\prime}}^2/2\sigma^2}\rho_{\mathrm{B}}(\varphi^{\prime}) \\
&=\begin{pmatrix}
R & i\sqrt{RD}e^{-i\varphi-\sigma^2/2}\\
-i\sqrt{RD}e^{i\varphi-\sigma^2/2} & D
\end{pmatrix}.
\end{split}
\end{align} 
The effect of the voltage fluctuations on the Bloch vector of the teleported state is to shrink the $x$ and $y$ components by a factor $e^{-\sigma^2/2}$
\begin{equation}\label{eq:rPrime}
\vecB{r}^{\prime} = 
\begin{pmatrix}
e^{-\sigma^2/2}r_x\\
e^{-\sigma^2/2}r_y\\
r_z
\end{pmatrix},
\end{equation} 
and therefore corresponds to phase-damping \cite{MikeAndIke}.
In order to quantify the effect on  the teleportation protocol, we calculate the teleportation fidelity $\telF$, which is the fidelity \cite{JozsaFidelity} between the input state and the output state averaged over all input states, with perfect teleportation  resulting in $\telF = 1$. The fidelity between two qubits with Bloch vectors $\vecB{r}$ and $\vecB{r}'$ is given by \cite{JozsaFidelity} 
\begin{equation}\label{eq:Fidelity}
\mathcal{F}=\frac{1}{2}\left\lbrace 1 + \vecB{r}\cdot\vecB{r}' + \left[\left(1-\norm{\vecB{r}}^2\right)\left(1-\norm{\vecB{r}'}^2\right)\right]^{1/2}\right\rbrace.
\end{equation}
The Bloch vector for the input state, given by the prepared state in the absence of decoherence, is 
\begin{equation}\label{eq:r}
\vecB{r} = \begin{pmatrix}
2\sqrt{RD}\sin\varphi\\
-2\sqrt{RD}\cos\varphi\\
R-D
\end{pmatrix}.
\end{equation}
We note that the state that reaches Alice may differ from the input state, as it has already been subject to fluctuating phases while propagating towards Alice.
Equations \eqref{eq:rPrime}, \eqref{eq:Fidelity} and \eqref{eq:r} then gives the fidelity for each input state as
\begin{equation}
\mathcal{F} = \frac{1+4e^{-\sigma^2/2}RD + (R-D)^2}{2}.
\end{equation}
Using $D=1-R$, and averaging the input state over the Bloch sphere results in a teleportation fidelity of
\begin{equation}\label{eq:SAWFidelity}
\telF = \frac{2 + e^{-\sigma^2/2}}{3}.
\end{equation}
If we consider a classical teleportation scheme where Alice measures the occupation of the A$^{\prime}$ modes and sends the result to Bob for him to prepare as the output state, the fidelity will be 2/3. This value is the maximum value that can be achieved using classical strategies~\cite{FidelityLimit}. Therefore, a fidelity above 2/3 implies that quantum resources are utilized \cite{NaturePhotonics}.  
Equation~\eqref{eq:SAWFidelity} shows that for $\sigma \to 0$, we recover the idealized picture $\telF \to 1$, while $\telF$ approaches the classical limit as $\sigma \to \infty$. In this limit, we lose any information about $\varphi$, just like in the classical scheme.

\subsection{Levitons and chiral edge states}\label{sec:Levitons}
\begin{figure}
\centering
\includegraphics[width=.9\columnwidth]{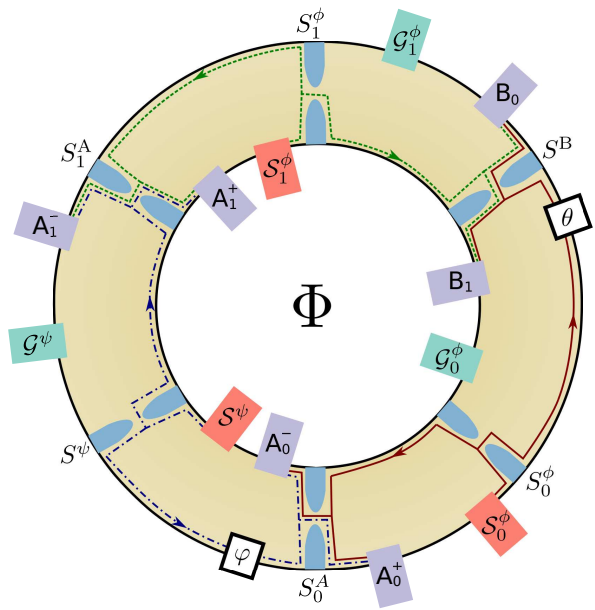}
\caption[The suggested Corbino setup]{Illustration of the experimental setup using levitons in chiral edge states, using a Corbino disk geometry. Electrons are generated at the sources (red) and travel along the edges of the disk, with the possible trajectories indicated by lines. The QPCs (blue) act as beamsplitters and facilitate scattering between the two edges. Phase differences $\varphi$ and $\theta$ are introduced along two of the paths and their sum, which is what the observables of interest depend on, can be tuned by a magnetic flux $\Phi$. Current measurements are performed at six different detectors (purple). We note that the same physical contacts may serve as the grounded contacts (green) and the detectors (purple).}\label{fig:Corbino}
\end{figure} 

Here we discuss a second experimental implementation, based on levitons traveling in chiral edge states. Such edge states do not host single modes, as considered above, but rather a continuum of modes that can be labeled by their energy. At zero temperature, all modes with energy below the chemical potential are occupied and represent the Fermi sea. Levitons are single-electron excitations above the Fermi sea generated by applying Lorentzian voltage pulses to metallic contacts \cite{LevitovMatPhys, Levitov}. Since transport occurs in a restricted range of energies around the chemical potential, we can assume all electrons to travel at the same velocity and the scattering induced by the beam-splitters to be energy independent \cite{Blanter}. The considered setup, which relies on a Corbino disk geometry, is sketched in Fig.~\ref{fig:Corbino}. Since single-electron detection is challenging for this type of setup, we provide a way to perform the state tomography measurements by measuring direct currents and zero frequency cross-correlators up to (and including) order three. This is enabled by a correspondence between these correlators and the observables in the idealized picture, allowing us to determine average values of these observables without access to single-shot detection. In order to generate the currents, the teleportation experiment will be repeated by periodically injecting levitons at each of the sources. The relevant quantities can be calculated using Floquet scattering theory \cite{Moskalets}. Quantum point contacts (QPC) act as beamsplitters and a magnetic flux $\Phi$ can be used to tune the sum $\varphi+\theta$, replacing the phase shifters required for the preparation and tomography steps. The geometry in Fig.\ \ref{fig:Corbino} is analogous to the one for the $N$-particle Aharonov-Bohm effect with $N=3$ \cite{2PartAB,3PartHBT}, where the $N$-th order zero frequency cross correlators oscillate as a function of $\Phi$ because the combined paths of $N$ electrons enclose the flux. 

\begin{table*}[t!]
	\begin{ruledtabular}
		\begin{tabular}{ccccc}
			Quantity & $D^{\prime} = 1/2$, $\theta = \pi/2$ & $D^{\prime} = 1/2$, $\theta = 0$ & $D^{\prime} = 1$, $\theta = 0$&$T$ dep.\\
			\hline
			$I_{\mathrm{A}_0^+}$, $I_{\mathrm{A}_0^-}$ & $\frac{e}{\Tau}\left(\frac{1}{4}+\frac{R}{2}\right)$ & $\frac{e}{\Tau}\left(\frac{1}{4}+\frac{R}{2}\right)$  & $\frac{e}{\Tau}\left(\frac{1}{4}+\frac{R}{2}\right)$ & 1\\
			$I_{\mathrm{A}_1^+}$, $I_{\mathrm{A}_1^-}$  & $\frac{e}{\Tau}\left(\frac{1}{4}+\frac{D}{2}\right)$ & $\frac{e}{\Tau}\left(\frac{1}{4}+\frac{D}{2}\right)$  & $\frac{e}{\Tau}\left(\frac{1}{4}+\frac{D}{2}\right)$ & 1\\
			$I_{\mathrm{B}_0}$, $I_{\mathrm{B}_1}$ & $\frac{e}{2\Tau}$ & $\frac{e}{2\Tau}$ &$\frac{e}{2\Tau}$ & 1\\
			$\mathcal{P}_{\mathrm{A}_0^{\pm}\mathrm{A}_1^{\pm}}$ &$ -\frac{e^2RD}{4\Tau}$&$ -\frac{e^2RD}{4\Tau}$&$ -\frac{e^2RD}{4\Tau}$&$F(T)$\\
			$\mathcal{P}_{\mathrm{A}_0^{+}\mathrm{B}_0}$, $\mathcal{P}_{\mathrm{A}_1^{+}\mathrm{B}_1}$ &$-\frac{e^2}{16\Tau}$&$-\frac{e^2}{16\Tau}$ & $-\frac{e^2}{8\Tau}$ &$F(T)$\\
			$\mathcal{P}_{\mathrm{A}_0^{+}\mathrm{A}_0^{-}}$, $\mathcal{P}_{\mathrm{A}_1^{+}\mathrm{A}_1^{-}}$ & $-\frac{e^2}{\Tau}\left(\frac{1}{16} - \frac{RD}{4}\right)$ & $-\frac{e^2}{\Tau}\left(\frac{1}{16} - \frac{RD}{4}\right)$ & $-\frac{e^2}{\Tau}\left(\frac{1}{16} - \frac{RD}{4}\right)$ & $F(T)$\\
			$\mathcal{P}_{\mathrm{A}_0^{+}\mathrm{B}_1}$, $\mathcal{P}_{\mathrm{A}_1^{+}\mathrm{B}_0}$& $-\frac{e^2}{16\Tau}$ & $-\frac{e^2}{16\Tau}$ & 0 &$F(T)$\\
			$\mathcal{Q}_{\mathrm{A}_0^{+}\mathrm{A}_1^{+}\mathrm{B}_0}$& $\frac{e^3}{\Tau}\frac{\sqrt{RD}\sin \varphi}{16}$ &$-\frac{e^3}{\Tau}\frac{\sqrt{RD}\cos \varphi}{16}$ &0 &$A(T)$\\
			$\mathcal{Q}_{\mathrm{A}_0^{+}\mathrm{A}_1^{+}\mathrm{B}_1}$&$-\frac{e^3}{\Tau}\frac{\sqrt{RD}\sin \varphi}{16}$  &$\frac{e^3}{\Tau}\frac{\sqrt{RD}\cos \varphi}{16}$ & 0&$A(T)$\\
			$\mathcal{Q}_{\mathrm{A}_0^{+}\mathrm{A}_0^{-}\mathrm{A}_1^+}$& $\frac{e^3}{\Tau}\frac{RD(R-D)}{8}$ & $\frac{e^3}{\Tau}\frac{RD(R-D)}{8}$ & $\frac{e^3}{\Tau}\frac{RD(R-D)}{8}$&$A(T)$\\
			$\mathcal{Q}_{\mathrm{A}_0^{+}\mathrm{A}_1^{+}\mathrm{A}_1^-}$& $\frac{e^3}{\Tau}\frac{RD(D-R)}{8}$ &$\frac{e^3}{\Tau}\frac{RD(D-R)}{8}$  &$\frac{e^3}{\Tau}\frac{RD(D-R)}{8}$ &$A(T)$\\
		\end{tabular}
	\end{ruledtabular}
	\caption{\label{tab:correlators}Expressions for the currents and correlators that are needed to demonstrate teleportation in the leviton architecture. The expressions are given for the three different tomography measurements required to determine Bob's state. The last column gives the temperature dependence for each quantity.}
\end{table*}

\subsubsection{Periodic driving}
 Since we are interested in periodically generating levitons, the  voltage applied to the source contacts is described by a train of Lorentizan pulses of width $\Gamma$ separated by the period $\Tau$
\begin{equation}
eV(t) = \sum_{j=-\infty}^{\infty} \frac{2\hbar\Gamma}{(t-j\Tau)^2 + \Gamma^2}.
\end{equation}
The electrons in the contact will pick up a time-dependent phase $\phi(t) = -\frac{1}{\hbar}\int_{-\infty}^{t}\diffD t^{\prime}eV(t^{\prime})$. This can be interpreted as the electrons exchanging energy quanta $\hbar\Omega$ with the voltage drive, where $\Omega=2\pi/\Tau$. The probability amplitude for exchanging $n$ quanta is given by the Fourier coefficients of the phase factor \cite{Moskalets}
\begin{equation}
S(n) = \frac{1}{\Tau}\int_0^{\Tau}\diffD t e^{in\Omega t}e^{i\phi(t)}.
\end{equation}
The resulting amplitudes for Lorentzian pulses are
\begin{equation}\label{eq:LevitonS}
 S(n) = 
\begin{cases}
-2e^{-n\Omega\Gamma}\sinh(\Omega\Gamma) & n>0,\\
e^{-\Omega\Gamma} & n=0,\\
0 & n<0.
\end{cases}
\end{equation}
 The excitations that are created by the voltage pulses are single-particle excitations called levitons. At zero temperature, they can be described by the following annihilation operator \cite{Levitov}
\begin{equation}\label{eq:LevitovA}
A_{\alpha} = \sqrt{2\Gamma}\sum_{E>\mu}e^{(it_{\alpha} -\Gamma)E/\hbar}a_{\alpha}(E),
\end{equation}
where $t_{\alpha}$ is the time at which the leviton is created and $\mu$ denotes the chemical potential of the contact.

The Floquet scattering matrix connects incoming states with energy $E$ in lead $\beta$ to outgoing states in lead $\alpha$ with energy $E_n = E + n\hbar\Omega$ \cite{Moskalets}.   For our system it is given by 
\begin{equation}\label{eq::SeparatedScattMat}
S_F(E_{n},E)_{\alpha\beta}=S_{\alpha\beta}S_{\beta}(n),
\end{equation}
where $S_{\beta}(n) = S(n)$ if $\beta$ is a source contact and $S_{\beta}(n) = \delta_{n,0}$ for the grounded contacts. The scattering matrix $S_{\alpha\beta}$ describes the setup presented in Sec.~\ref{sec:IdealSetup}, see Eq. \eqref{eq:totalScatteringMatrix}.

\subsubsection{Observables and zero temperature results}
In order to demonstrate that teleportation can be achieved in this architecture, we express the desired observables in terms of the zero frequency correlators
\begin{align}\label{eq:correlatorIntegrals}
\begin{split}
I_{\alpha} &= \frac{1}{\Tau}\int_0^{\Tau}\diffD t \langle \hat{I}_{\alpha}(t) \rangle, \\
\mathcal{P}_{\alpha\beta} &= \frac{1}{\Tau}\int_0^{\Tau}\diffD t \int_{-\infty}^{\infty}\diffD \tau \langle \Delta \hat{I}_{\alpha}(t)\Delta \hat{I}_{\beta}(t+\tau) \rangle,\\
\mathcal{Q}_{\alpha\beta\gamma} &= \frac{1}{\Tau}\int_0^{\Tau}\diffD t\int_{-\infty}^{\infty}\diffD \tau_{\beta}\int_{-\infty}^{\infty}\diffD \tau_{\gamma} \langle \Delta \hat{I}_{\alpha}(t)\\
&\times\Delta \hat{I}_{\beta}(t+\tau_{\beta})\Delta \hat{I}_{\gamma}(t+\tau_{\gamma}) \rangle,
\end{split}
\end{align} 
 where $\hat{I}_{\alpha}(t)$ is the current operator in lead $\alpha$ and $\Delta \hat{I}_{\alpha}(t)=\hat{I}_{\alpha}(t)-\langle \hat{I}_{\alpha}(t)\rangle$. The averages are calculated by assuming that the electrons in each of the contacts are distributed according to the Fermi distribution 
 \begin{equation}
f_{\alpha}(E) = \frac{1}{e^{E/\kB T}+1}.
\end{equation} We assume that all the contacts are kept at the same temperature $T$, and at vanishing chemical potential. In this section we will treat the case $T=0$, while the picture at finite temperature is presented in the following section.

The number of excess electrons in lead $\alpha$ within one period is given by
\begin{equation}\label{eq:ExecessElectrons}
N_{\alpha}=\frac{1}{e}\int_0^{\Tau}\diffD t \hat{I}_{\alpha}(t).
\end{equation} 
Under the assumption that only electrons within the same period are correlated, the higher moments of this operator are determined by the zero frequency current correlators.
The explicit form of the correspondence is, defining $\Delta N_\alpha=N_{\alpha}-\langle N_{\alpha}\rangle$
\begin{align}\label{eq:Correspondence}
\begin{split}
&\langle N_{\alpha}\rangle = \frac{\Tau}{e}I_{\alpha},\\
&\langle \Delta N_{\alpha} \Delta N_{\beta} \rangle = \frac{\Tau}{e^2} \mathcal{P}_{\alpha\beta},\\
&\langle \Delta N_{\alpha} \Delta N_{\beta} \Delta N_{\gamma} \rangle = \frac{\Tau}{e^3}\mathcal{Q}_{\alpha\beta\gamma},
\end{split}
\end{align}
which is shown in App.~\ref{App:Correspondence}. These expressions are expected to be valid if the voltage pulses are well separated so that consecutive wave packets have small overlap. Note that connecting state tomography observables to measurements of current correlators has been suggested previously, see e.g. Ref.~\cite{StateTomoPeter,jullien:2014}.

To make the correspondence between the Floquet correlators and tomography observables concrete, we write out Bob's Bloch vector in terms of current correlators using Eqs.~\eqref{eq:TomoExpr} and \eqref{eq:Correspondence}
\begin{equation}\label{eq:CorrespondanceExact}
r_i^{\prime} = \frac{J}{K},
\end{equation}
where we have defined
\begin{widetext}
\begin{align}\label{eq:LevitonObservables}
\begin{split}
J&=\frac{\Tau}{e^3}\biggl(\mathcal{Q}_{\mathrm{A}_0^+\mathrm{A}_1^+\mathrm{B}_0}-\mathcal{Q}_{\mathrm{A}_0^+\mathrm{A}_1^+\mathrm{B}_1}\biggr)+ \frac{\Tau^2}{e^3}\biggl[\mathcal{P}_{\mathrm{A}_0^+\mathrm{A}_1^+}(I_{\mathrm{B}_0}-I_{\mathrm{B}_1})+ I_{\mathrm{A}_1^+}(\mathcal{P}_{\mathrm{A}_0^+\mathrm{B}_0}-\mathcal{P}_{\mathrm{A}_0^+\mathrm{B}_1})+I_{\mathrm{A}_0^+}(\mathcal{P}_{\mathrm{A}_1^+\mathrm{B}_0}-\mathcal{P}_{\mathrm{A}_1^+\mathrm{B}_1})\biggr] \\
&+ \frac{\Tau^3}{e^3}I_{\mathrm{A}_0^+}I_{\mathrm{A}_1^+}\biggr(I_{\mathrm{B}_0}-I_{\mathrm{B}_1}\biggr),\\
K&=\frac{\Tau^2}{e^2}I_{\mathrm{A}_0^+}I_{\mathrm{A}_1^+}\biggl(1-\frac{\Tau}{e}(I_{\mathrm{A}_0^-}+I_{\mathrm{A}_1^-})\biggr)-\frac{\Tau^2}{e^3}\biggl[I_{\mathrm{A}_0^+}(\mathcal{P}_{\mathrm{A}_0^-\mathrm{A}_1^+}+\mathcal{P}_{\mathrm{A}_1^+\mathrm{A}_1^-})
+I_{\mathrm{A}_1^+}(\mathcal{P}_{\mathrm{A}_0^+\mathrm{A}_0^-}+\mathcal{P}_{\mathrm{A}_0^+\mathrm{A}_1^-})\biggr]\\ 
&-\frac{\Tau}{e^3}\biggl(\mathcal{Q}_{\mathrm{A}_0^+\mathrm{A}_1^+\mathrm{A}_0^{-}}+\mathcal{Q}_{\mathrm{A}_0^+\mathrm{A}_1^+\mathrm{A}_1^-}\biggr).
\end{split}
\end{align}
\end{widetext}

The zero temperature expressions for the correlators involved in Eq. \eqref{eq:LevitonObservables}, for each of the three settings of the tomography setup, can be found in Tab.~\ref{tab:correlators} and we recover $\vecB{r}^{\prime}=\vecB{r}$. Note that $K$ gives the value of $p(+,+)$, equal to $1/16$ at zero temperature. This demonstrates that teleportation with levitons is achieved at zero temperature. 

\subsubsection{Finite temperature results}
At finite temperatures, thermal excitations complicate the simple picture of teleportation that has been presented so far. These additional excitations imply that we no longer have a qubit system but a more complicated many-body state.
The POVM that was defined for Alice's electron detection is no longer appropriate to describe the situation, since it assumes that there are no more than three electrons present. Furthermore, since each edge state hosts a continuum of modes, $N_i$ can no longer be treated as an operator with eigenvalues 0 and 1.
 Nevertheless, the definition in Eq.~\eqref{eq:CorrespondanceExact} can be extended to finite temperatures. At low temperatures, where the number of additional excitations is small, we find the situation to be well described by noisy teleportation.  
 
  \begin{figure}[t!]
  	
  	\includegraphics[width=\columnwidth]{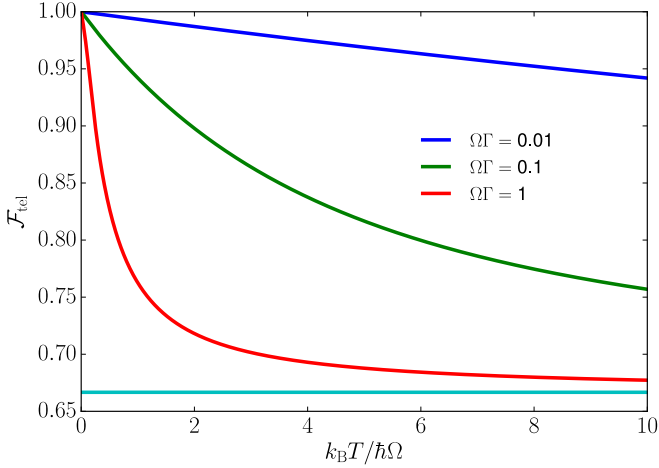}
  	\caption{\label{fig:TempDependence}Teleportation fidelity for the ++ outcome as a function of temperature for different leviton widths. The horizontal line denotes the classical limit of $\telF = 2/3$.}
  \end{figure}
 
For  finite temperatures, the current is unaffected, while the second and third order correlators each pick up a temperature dependent factor
\begin{align}
\begin{split}
&\mathcal{P}_{\alpha\beta}(T)=F(T)\mathcal{P}_{\alpha\beta}(0),\\
&\mathcal{Q}_{\alpha\beta\gamma}(T) = A(T)\mathcal{Q}_{\alpha\beta\gamma}(0),
\end{split}
\end{align}
where
\begin{align}
\begin{split}
F(T)&=\sum_{n=1}^{\infty}n\left(\coth \frac{n\hbar\Omega}{2k_{\mathrm{B}}T}- 2\frac{k_{\mathrm{B}}T}{n\hbar\Omega}\right)\left|S(n)\right|^2,\\
A(T)&=\sum_{n=1}^{\infty}n\biggl(\coth^2\frac{n\hbar\Omega}{2\kB T} + \frac{1}{2}\csch^2\frac{n\hbar\Omega}{2\kB T} \\
&- 3\frac{\kB T}{n\hbar\Omega}\coth\frac{n\hbar\Omega}{2\kB T}\biggr)\left|S(n)\right|^2,
\end{split}
\end{align}
with $S(n)$ given in Eq.~\eqref{eq:LevitonS}.
Using the above expressions to calculate the right-hand-side of Eq.~\eqref{eq:CorrespondanceExact} at finite temperatures results in
\begin{equation}\label{eq:BlochTransf}
\vecB{r}^{\prime}  = \begin{pmatrix}
q(T)r_x\\
q(T)r_y\\
r_z
\end{pmatrix},\hspace{1.5cm} q(T)=\frac{A(T)}{F(T)}.
\end{equation}
As $q(T)\leq1$, this is equivalent to teleportation affected by phase damping, just as in the SAW architecture when voltage fluctuations are included. For the input state $\boldsymbol{r}$, we again consider the state that is prepared in the absence of environmental effects, i.e., at zero temperature. At finite temperatures, the prepared state is a mixed state, which can be interpreted as a state preparation that may fail, reducing the teleportation fidelity.
The teleportation fidelity for levitons is then given by
\begin{equation}\label{eq:LevitonFidelity}
\telF = \frac{2 + q(T)}{3}.
\end{equation}
Figure \ref{fig:TempDependence} shows $\telF$ as a function of temperature for different leviton widths. As the temperature is increased the fidelity decreases and approaches the classical limit. This happens faster for broad levitons because a narrow leviton has a wider energy spectrum than a broad leviton. It will thus stand out more against the thermal excitations, which are relevant on a scale of $\kB T$ around the Fermi energy.

\section{Conclusions}\label{sec:conclusions}
We have theoretically demonstrated a  scheme for quantum teleportation of flying single-electron qubits. The efficiency of the scheme was studied in an idealized scenario, where we considered quantum state tomography of Bob's post-measurement state. The scheme is successful 25\% or 12.5\% of the time, depending on the presence or absence of a feed-forward scheme. We considered two experimental implementations based on SAWs and levitons in chiral edge states respectively. Single-electron detection is available for the SAW architecture, while it is currently not available for levitons. For the SAW approach we studied the effect of decoherence due to voltage fluctuations.  For the leviton approach, we showed how state tomography can be performed using low-frequency current correlators and we considered the effect of finite temperatures. In both implementations, the effect of the environment is captured well by phase damping. As charge couples to the environment through the Coulomb interaction, environmental effects may be comparably large when performing teleportation with electrons. However, voltage gates can be used to influence the Coulomb interaction, providing an avenue for mitigating environmental effects \cite{huynh:2012}.

Promising future avenues include the implementation of other protocols from quantum information using single-electron states, as well as an investigation of more efficient teleportation protocols by relaxing the particle number superselection \cite{FriisTel}, which could be achieved using superconductors.

\begin{acknowledgments}
 We acknowledge fruitful discussions with Nicolai Friis. This work was supported by the Swedish Research Council. E.O. acknowledges support from Jan Marcus Dahlstr\"om through the Swedish Foundations' Starting Grant funded by Olle Engkvists Stiftelse. N.B. acknowledges funding from the Swiss National Science Foundation through the starting grant DIAQ and the NCCR QSIT. P.P.P. acknowledges funding from the European Union's Horizon 2020 research and innovation programme under the Marie Sk{\l}odowska-Curie Grant Agreement No. 796700.
\end{acknowledgments}

\appendix
\section{Full scattering matrix}\label{App:Full}
The total scattering matrix for the idealized teleportation setup, including the state tomography part of the setup reads
\begin{widetext}
	\begin{equation}\label{eq:totalScatteringMatrix}
	\begin{pmatrix}
	a_{\mathrm{A}_0^+}\\
	a_{\mathrm{A}_0^-}\\
	a_{\mathrm{A}_1^+}\\
	a_{\mathrm{A}_1^-}\\
	a_{\mathrm{B}_0}\\
	a_{\mathrm{B}_1}
	\end{pmatrix}
	=
	\frac{1}{\sqrt{2}}
	\begin{pmatrix}
	\frac{-1}{\sqrt{2}} & \frac{i}{\sqrt{2}} &  0 & 0 & i\sqrt{R}e^{-i\varphi} & \sqrt{D}e^{-i\varphi}\\
	\frac{i}{\sqrt{2}} &  \frac{1}{\sqrt{2}} &  0 & 0 & -\sqrt{R}e^{-i\varphi} & i\sqrt{D}e^{-i\varphi}\\
	0 &0 &\frac{-1}{\sqrt{2}} & \frac{i}{\sqrt{2}} & \sqrt{D} & i\sqrt{R} \\
	0 &0 &\frac{i}{\sqrt{2}} & \frac{1}{\sqrt{2}} & i\sqrt{D} & -\sqrt{R} \\
	\sqrt{D^{\prime}}e^{-i\theta} & i\sqrt{D^{\prime}}e^{-i\theta} & -i\sqrt{R^{\prime}} & \sqrt{R^{\prime}} & 0 & 0\\
	-i\sqrt{R^{\prime}}e^{-i\theta} & \sqrt{R^{\prime}}e^{-i\theta} & \sqrt{D^{\prime}} & i\sqrt{D^{\prime}} & 0 & 0\\
	\end{pmatrix}
	\begin{pmatrix}
	a_{\mathcal{S}_0^{\phi}}\\
	a_{\mathcal{G}_0^{\phi}}\\
	a_{\mathcal{S}_1^{\phi}}\\
	a_{\mathcal{G}_1^{\phi}}\\
	a_{\mathcal{S}^{\psi}}\\
	a_{\mathcal{G}^{\psi}}
	\end{pmatrix}.
	\end{equation}

\section{States and operators in second quantization}\label{App:ProjectionsSQ}
The Bell states $\lbrace \myKet{\Uppsi^\pm_{\mathrm{AB}^{\prime}}},\myKet{\Upphi^\pm_{\mathrm{AB}^{\prime}}}\rbrace$ can be written in terms of creation operators as
\begin{align}\label{eq:BellStates}
\begin{split}
&\myKet{\Uppsi^+_{\mathrm{A}\mathrm{B}^{\prime}}} =\frac{1}{\sqrt{2}}(\aDagg{\mathrm{A}_0}\aDagg{\mathrm{B}^{\prime}_1}+\aDagg{\mathrm{A}_1}\aDagg{\mathrm{B}^{\prime}_0})\ket{\Omega},\hspace{2cm}\myKet{\Uppsi^-_{\mathrm{A}\mathrm{B}^{\prime}}} = 
\frac{1}{\sqrt{2}}(\aDagg{\mathrm{A}_0}\aDagg{\mathrm{B}^{\prime}_1}-\aDagg{\mathrm{A}_1}\aDagg{\mathrm{B}^{\prime}_0})\ket{\Omega},\\
&\myKet{\Upphi^+_{\mathrm{A}\mathrm{B}^{\prime}}} =\frac{1}{\sqrt{2}}(\aDagg{\mathrm{A}_0}\aDagg{\mathrm{B}^{\prime}_0}+\aDagg{\mathrm{A}_1}\aDagg{\mathrm{B}^{\prime}_1})\ket{\Omega},\hspace{2cm}\myKet{\Upphi^-_{\mathrm{A}\mathrm{B}^{\prime}}} =\frac{1}{\sqrt{2}} (\aDagg{\mathrm{A}_0}\aDagg{\mathrm{B}^{\prime}_0}-\aDagg{\mathrm{A}_1}\aDagg{\mathrm{B}^{\prime}_1})\ket{\Omega}.
\end{split}
\end{align}

Next we give the second quantization representation of the states appearing in Eq.~\eqref{eq:totalProjectedState}
\begin{align}\label{eq:ProjectionsSQ}
\begin{split}
&\myKet{\Uppsi^-_{\mathrm{A}^{\prime}\mathrm{A}},\psi_{\mathrm{B}^{\prime}}} = \frac{1}{\sqrt{2}}(\aDagg{\mathrm{A}^{\prime}_0}\aDagg{\mathrm{A}_1}-\aDagg{\mathrm{A}^{\prime}_1}\aDagg{\mathrm{A}_0})(i\sqrt{R}e^{-i\varphi}a^{\dagger}_{\mathrm{B}^{\prime}_0} + \sqrt{D}a^{\dagger}_{\mathrm{B}_{1}^{\prime}})\ket{\Omega},\\
&\myKet{\Uppsi^+_{\mathrm{A}^{\prime}\mathrm{A}},\sigma_z\psi_{\mathrm{B}^{\prime}}} =\frac{1}{\sqrt{2}}(\aDagg{\mathrm{A}^{\prime}_0}\aDagg{\mathrm{A}_1}+\aDagg{\mathrm{A}^{\prime}_1}\aDagg{\mathrm{A}_0})(i\sqrt{R}e^{-i\varphi}a^{\dagger}_{\mathrm{B}_{0}^{\prime}} - \sqrt{D}a^{\dagger}_{\mathrm{B}_{1}^{\prime}})\ket{\Omega},\\
&\myKet{\Upphi^+_{\mathrm{A}^{\prime}\mathrm{A}}, \sigma_x\psi_{\mathrm{B}^{\prime}}} = \frac{1}{\sqrt{2}}(\aDagg{\mathrm{A}^{\prime}_0}\aDagg{\mathrm{A}_0}+\aDagg{\mathrm{A}^{\prime}_1}\aDagg{\mathrm{A}_1})(\sqrt{D}a^{\dagger}_{\mathrm{B}_{0}^{\prime}} + i\sqrt{R}e^{-i\varphi}a^{\dagger}_{\mathrm{B}_{1}^{\prime}})\ket{\Omega} ,\\
&\myKet{\Upphi^-_{\mathrm{A}^{\prime}\mathrm{A}}, i\sigma_y\psi_{\mathrm{B}^{\prime}}} =\frac{1}{\sqrt{2}} (\aDagg{\mathrm{A}^{\prime}_0}\aDagg{\mathrm{A}_0}-\aDagg{\mathrm{A}^{\prime}_1}\aDagg{\mathrm{A}_1})(\sqrt{D}a^{\dagger}_{\mathrm{B}_{0}^{\prime}} - i\sqrt{R}e^{-i\varphi}a^{\dagger}_{\mathrm{B}_{1}^{\prime}})\ket{\Omega}.
\end{split}
\end{align}

The explicit form of the $\ket{R}$ state introduced in Eq.~\eqref{eq:finalState} reads
	\begin{align}\label{eq:finalStateRemainder}
	\begin{split}
	\ket{R} &= \frac{1}{\sqrt{3}}\biggl\lbrace 
	- \frac{i\sqrt{R}e^{-i\varphi}}{\sqrt{2}}\left(i\aDagg{\mathrm{A}^+_0}\aDagg{\mathrm{A}^-_0}\aDagg{\mathrm{A}^+_1} + \aDagg{\mathrm{A}^+_0}\aDagg{\mathrm{A}^-_0}\aDagg{\mathrm{A}^-_1}\right) +\frac{\sqrt{D}}{\sqrt{2}}\left(i\aDagg{\mathrm{A}^+_0}\aDagg{\mathrm{A}^+_1}\aDagg{\mathrm{A}^-_1} + \aDagg{\mathrm{A}^-_0}\aDagg{\mathrm{A}^+_1}\aDagg{\mathrm{A}^-_1} \right) \\
	& + \frac{1}{\sqrt{2}}\Bigl[i\sqrt{R}e^{-i\varphi}\left(\aDagg{\mathrm{A}^+_0} + i\aDagg{\mathrm{A}^-_0}\right)+  \sqrt{D}\left(\aDagg{\mathrm{A}^+_1}+i\aDagg{\mathrm{A}^-_1}\right)\Bigr]\aDagg{\mathrm{B}_0^{\prime}}\aDagg{\mathrm{B}_1^{\prime}}-\sqrt{R}e^{-i\varphi}\aDagg{\mathrm{A}^+_0}\aDagg{\mathrm{A}^-_0}\aDagg{\mathrm{B}_1^{\prime}} - i\sqrt{D}\aDagg{\mathrm{A}^+_1}\aDagg{\mathrm{A}^-_1}\aDagg{\mathrm{B}_0^{\prime}} \biggr\rbrace \ket{\Omega}. 
	\end{split}
	\end{align}
\end{widetext}
As mentioned in Sec.~\ref{sec:electronDetection} this part of $\ket{\Psi}$ is not useful for teleportation purposes with our setup. The terms in $\ket{R}$  do not generate measurement outcomes where Bob is left with a dual-rail qubit when Alice has performed her measurement. Some terms correspond to having all particles at Alice's location, which defeats the purpose of trying to send a qubit state to Bob. Others have two particles at the same A$_{j}$, which means that Bob cannot have a superposition of his two modes. In the remaining cases Bob ends up with two of the particles, which means that both of Bob's modes will be occupied, and he can therefore not have a dual-rail qubit.

The Pauli matrices in the dual-rail qubit space in terms of $\rmB^{\prime}$ modes read
\begin{align}
\begin{split}
\sigma_x &= \aDagg{\rmB_1^{\prime}}a_{\rmB_0^{\prime}} + \aDagg{\rmB_0^{\prime}}a_{\rmB_1^{\prime}},\\
\sigma_y &= i(\aDagg{\rmB_1^{\prime}}a_{\rmB_0^{\prime}} - \aDagg{\rmB_0^{\prime}}a_{\rmB_1^{\prime}}),\\
\sigma_z &= \aDagg{\rmB_0^{\prime}}a_{\rmB_0^{\prime}} - \aDagg{\rmB_1^{\prime}}a_{\rmB_1^{\prime}}.
\end{split}
\end{align} 

\section{The POVM}\label{App:POVM}
A general POVM element describing the detection of the particle number of each mode at A is given by
\begin{equation}\label{eq:defPOVM}
E\left(j_{\mathrm{A}_0^+},j_{\mathrm{A}_0^-}, j_{\mathrm{A}_1^+},j_{\mathrm{A}_1^-}\right) = \prod_iN_i^{j_i}\left(I-N_i\right)^{(1-j_i)},
\end{equation}
where $j_i\in \lbrace 0,1\rbrace$ is the number of electrons detected in mode $i$, with $i \in \{\mathrm{A}_0^+,\mathrm{A}_0^-,\mathrm{A}_1^+,\mathrm{A}_1^-\}$. $N_i = \aDagg{i}a_i$ is the particle number operator for mode $i$. The factor $N_i^{j_i}\left(I-N_i\right)^{(1-j_i)}$ in the POVM will  project states onto the subspace with $j_i$ particles in mode $i$. For the POVM elements used in the main text we have $E(+,+) = E(1,0,1,0)$ etc.
That the operators defined by Eq.~\eqref{eq:defPOVM} are positive follows from the fact that they are products of operators with eigenvalues 0 and 1.  
A straightforward calculation shows that
\begin{equation}\label{eq:POVMIdentity}
\sum_X E(X) = I,
\end{equation}
with $X \in \{(j_{\mathrm{A}_0^+},j_{\mathrm{A}_0^-},j_{\mathrm{A}_1^+},j_{\mathrm{A}_1^-})|j_i\in \{0,1\}\}$.

We can write the POVM elements in the dual-rail qubit space by using the projection operator
\begin{equation}
P_{\mathrm{drq}} = \sum_j \myKet{j}\myBra{j}
\end{equation}
where the $\myKet{j}$ states are the Bell-states from Eq.~\eqref{eq:BellStates}. Defining $E_{\mathrm{drq}}(X)$ as
\begin{equation}
E_{\mathrm{drq}}(X)  = P_{\mathrm{drq}} E(X) P_{\mathrm{drq}},
\end{equation} 
we find that the projections of the POVM elements corresponding to finding two electrons at A are
\begin{align}
\begin{split}
E_{\mathrm{drq}}(1,0,1,0) &= E_{\mathrm{drq}}(0,1,0,1) = \frac{1}{2}\myKet{\Uppsi^-_{\mathrm{A}^{\prime}\mathrm{A}}} \myBra{\Uppsi^-_{\mathrm{A}^{\prime}\mathrm{A}}},\\
E_{\mathrm{drq}}(1,0,0,1) &= E_{\mathrm{drq}}(0,1,1,0) = \frac{1}{2} \myKet{\Uppsi^+_{\mathrm{A}^{\prime}\mathrm{A}}} \myBra{\Uppsi^+_{\mathrm{A}^{\prime}\mathrm{A}}},\\
 E_{\mathrm{drq}}(1,1,0,0) &= \frac{1}{2}(\myKet{\Upphi^+_{\mathrm{A}^{\prime}\mathrm{A}}}+\myKet{\Upphi^-_{\mathrm{A}^{\prime}\mathrm{A}}})(\myBra{\Upphi^+_{\mathrm{A}^{\prime}\mathrm{A}}}+\myBra{\Upphi^-_{\mathrm{A}^{\prime}\mathrm{A}}}),\\
E_{\mathrm{drq}}(0,0,1,1) &= \frac{1}{2}(\myKet{\Upphi^+_{\mathrm{A}^{\prime}\mathrm{A}}}-\myKet{\Upphi^-_{\mathrm{A}^{\prime}\mathrm{A}}})(\myBra{\Upphi^+_{\mathrm{A}^{\prime}\mathrm{A}}}-\myBra{\Upphi^-_{\mathrm{A}^{\prime}\mathrm{A}}}).
\end{split}
\end{align}

\section{Current and number operators}\label{App:Correspondence}
Here we prove the relations given in Eq.~\eqref{eq:Correspondence}.
From the definitions in Eqs.~\eqref{eq:correlatorIntegrals} and \eqref{eq:ExecessElectrons}, the expectation value for $N_{\alpha}$ is seen to be 
\begin{equation}
\langle N_{\alpha}\rangle = \frac{\Tau}{e}I_{\alpha}.
\end{equation}
With $\Delta N_{\alpha} = N_{\alpha}-\langle N_{\alpha} \rangle$, we find 
\begin{align}\label{eq:numberNoise}
\begin{split}
\langle \Delta N_{\alpha} \Delta N_{\beta} \rangle &= \frac{1}{e^2}\int_0^{\Tau}\diffD t \int_0^{\Tau}\diffD t^{\prime} 
\langle \Delta \hat{I}_{\alpha}(t)\Delta \hat{I}_{\beta}(t^{\prime}) \rangle.
\end{split}
\end{align} 
We now assume that $\Delta \hat{I}_{\alpha}(t)$ and $\Delta \hat{I}_{\beta}(t^{\prime})$ are uncorrelated when $t$ and $t^{\prime}$ lie in different periods, i.e.\ $\langle \Delta \hat{I}_{\alpha}(t)\Delta \hat{I}_{\beta}(t^{\prime})\rangle = 0$ for $\lfloor t/\Tau \rfloor \neq \lfloor t^{\prime}/\Tau \rfloor $, where $\lfloor x\rfloor$ is the floor function. This assumption is expected to be valid if the Lorentzian voltage pulses have small widths, such that subsequent pulses do not overlap significantly. We can then extend the integral over $t^{\prime}$ to range from $-\infty$ to $\infty$. A change of variables, $t^{\prime} = t+\tau$, results in
\begin{equation}
\langle \Delta N_{\alpha} \Delta N_{\beta} \rangle = \frac{\Tau}{e^2} \mathcal{P}_{\alpha\beta}.
\end{equation}

For $\langle \Delta N_{\alpha} \Delta N_{\beta} \Delta N_{\gamma} \rangle$ we get
\begin{align}
\begin{split}
\langle \Delta N_{\alpha} \Delta N_{\beta} \Delta N_{\gamma} \rangle &= \frac{1}{e^3}\int_0^{\Tau}\diffD t\int_{0}^{\Tau}\diffD \tau_{\beta}\int_{0}^{\Tau}\diffD \tau_{\gamma} \langle \Delta \hat{I}_{\alpha}(t)\\
&\times\Delta \hat{I}_{\beta}(\tau_{\beta})\Delta \hat{I}_{\gamma}(\tau_{\gamma}) \rangle.
\end{split}
\end{align}
Under the assumption that $\langle \Delta \hat{I}_{\alpha}(t)\Delta \hat{I}_{\beta}(t^{\prime})\Delta \hat{I}_{\gamma}(t^{\prime\prime}) \rangle$
vanishes unless $t$, $t^{\prime}$ and $t^{\prime\prime}$ all lie in the same period, we can again extend the integrals and change the integration variables to show
\begin{equation}
\langle \Delta N_{\alpha} \Delta N_{\beta} \Delta N_{\gamma} \rangle = \frac{\Tau}{e^3}\mathcal{Q}_{\alpha\beta\gamma}.
\end{equation}

\bibliography{tel_ref_paper}

\end{document}